\documentclass[aps,pra,showpacs,floatfix,twocolumn,superscriptaddress]{revtex4-1}
\usepackage{amssymb}
\usepackage{xr-hyper}
\usepackage{alclock}

\begin{document}

%%%%%%%%%%%%%%%%%%%%%%%%%%%%%%%%%%%%%%%%%%%%%%%%%%
% Header
%%%%%%%%%%%%%%%%%%%%%%%%%%%%%%%%%%%%%%%%%%%%%%%%%%

\title{Systematic uncertainty due to background-gas collisions in trapped-ion optical clocks}
\author{A. M. Hankin}
\altaffiliation{Present address: Honeywell Quantum Solutions, Broomfield, CO 80021}
\affiliation{\NIST}
\affiliation{\CU}
\author{E. R. Clements}
\affiliation{\NIST}
\affiliation{\CU}
\author{Y. Huang}
\affiliation{\WIPM}
\author{S. M. Brewer}
\affiliation{\NIST}
\affiliation{\CU}
\author{J.-S. Chen}
\altaffiliation{Present address: IonQ Inc., College Park, MD 20740}
\affiliation{\NIST}
\affiliation{\CU}
\author{C. W. Chou}
\affiliation{\NIST}
\author{D. B. Hume}
\affiliation{\NIST}
\author{D. R. Leibrandt}
\email{david.leibrandt@nist.gov}
\affiliation{\NIST}
\affiliation{\CU}
\date{\today}

\begin{abstract}
We describe a framework for calculating the frequency shift and uncertainty of trapped-ion optical atomic clocks caused by background-gas collisions, and apply this framework to an $\alp$ clock to enable a total fractional systematic uncertainty below $10^{-18}$.  For this clock, with 38(19)~nPa of room temperature H$_2$ background gas, we find that collisional heating generates a non-thermal distribution of motional states with a mean time-dilation shift of order $10^{-16}$ at the end of a 150~ms probe, which is not detected by sideband thermometry energy measurements.  However, the contribution of collisional heating to the spectroscopy signal is highly suppressed and we calculate the BGC shift to be $-0.6(2.4)\times 10^{-19}$, where the shift is due to collisional heating time-dilation and the uncertainty is dominated by the worst case $\pm \pi/2$ bound used for collisional phase shift of the $\alp$ superposition state.  We experimentally validate the framework and determine the background-gas pressure in situ using measurements of the rate of collisions that cause reordering of mixed-species ion pairs.
\end{abstract}

\maketitle

%%%%%%%%%%%%%%%%%%%%%%%%%%%%%%%%%%%%%%%%%%%%%%%%%%
% Introduction
%%%%%%%%%%%%%%%%%%%%%%%%%%%%%%%%%%%%%%%%%%%%%%%%%%

Background-gas collisions (BGCs) are a leading source of systematic frequency shifts and uncertainty in many of the most accurate microwave fountain~\cite{Li2011,Guena2012,Gibble2013}, optical lattice~\cite{McGrew2018,Bothwell2019,Alves2019}, and trapped-ion~\cite{Vutha2017,Huntemann2016,Dube2013,Chou2010} atomic clocks.  Both lattice and trapped-ion optical clocks, in particular, report fractional uncertainties at the low $10^{-18}$ level with BGCs contributing at the mid $10^{-19}$ level.  Further improvements to the systematic uncertainties of optical clocks will require either reduced background-gas (BG) pressures or more accurate evaluations of the BGC shift.

The frequency of atomic clocks is affected by BGCs in two distinct ways: they change the motional state distribution, which modifies the time-dilation (i.e., second-order Doppler) shift; and they change the phase of the atomic superposition state due to interactions during the collision.  Previous evaluations of the systematic uncertainty due to BGCs in trapped-ion optical clocks have focused on the collisional phase shift~\cite{Madej2004,Rosenband2008,Chou2010,Dube2013,Nisbet-Jones2016,Huntemann2016,Cao2017,Vutha2017} and implicitly assumed that the time-dilation shift due to collisional heating is captured by measurements of secular motion heating using sideband thermometry~\cite{Leibfried2003}.  This treatment might be adequate when there is continuous sympathetic cooling of the clock ions during the spectroscopy pulses~\cite{Chou2010}, but here we show that collisions with BG molecules in the absence of cooling generate a nonthermal distribution of motional Fock states with a mean energy that can be significantly larger than the Doppler limit. Sideband thermometry is known to underestimate the mean energy of motional state distributions with super-thermal tails~\cite{Chen2017}. Thus, time-dilation due to collisional heating may not have been fully included in previous clock evaluations.

At large separations $R$, the interaction potential between a polarizable neutral molecule and an ion is given by $V(R) = -\alpha Q^2 / (8 \pi \epsilon_0 R^4)$, independent of the internal state of the ion.  Here, $\alpha$ is the static polarizability of the molecule~\footnote{For H$_2$, the orientation averaged dc polarizability $\alpha = 0.8 \times 10^{-30}$~m$^3$~\cite{Landolt1951,Akindinova2009}.}, $Q$ is the charge of the ion, and $\epsilon_0$ is the vacuum permittivity \cite{Wineland1998}.  For small impact parameters, in the classical picture this potential leads to Langevin collision trajectories in which the molecule spirals in towards the ion.  At small separations $R$, the interaction potential is different for the ground and excited states of the clock transition, leading to a phase shift for Langevin collision events.  For large impact parameter events, the centrifugal barrier exceeds the kinetic energy and the short range potential is not sampled, so the phase shift is negligible.  However, for both Langevin and glancing collisions the BG molecule trajectory is deflected leading to heating of the ion.

In this paper, we present a theoretical calculation of the motional-state distribution of trapped ions that experience BGCs in an ultra-high vacuum (UHV) environment, with focus placed on $\alp$ quantum-logic clocks.  Next, we experimentally validate this calculation by measuring the reorder rate of a mixed species $\caalp$ ion pair as a function of the potential barrier between the two orders.  Our collision kinematics model agrees with the measurements over the range of experimentally accessible reorder energy barriers.  Finally, we evaluate the BGC shift for the $\mgalp$ clock described by \citet{Brewer2019} using a Monte-Carlo simulation of the clock interrogation including motional heating and phase shifts of the $\alp$ ion due to BGCs.

%%%%%%%%%%%%%%%%%%%%%%%%%%%%%%%%%%%%%%%%%%%%%%%%%%
% Collision kinematics model
%%%%%%%%%%%%%%%%%%%%%%%%%%%%%%%%%%%%%%%%%%%%%%%%%%

In the following, we consider collisions of H$_2$, the dominant gas in room temperature UHV systems, with $\mgalp$ or $\caalp$ ion pairs confined in linear Paul traps.  We are interested in the non-equilibrium distribution of ion motional states that occur when the motion is initially prepared near the three-dimensional motional ground state \cite{Chen2017} and evolves due to BGCs without sympathetic cooling during the clock interrogation time, which is short compared to the mean time between collisions with H$_2$.  Because the collision dynamics occur on a time scale ($\sim 10^{-12}$~s) short compared to the fastest time scale for ion motion ($\sim 10^{-8}$~s), it is sufficient to treat the collisions as leading to instantaneous changes to the velocity of one of the ions. Furthermore, we assume that in each collision event the BG molecule interacts with only one of the two ions.

Here, we summarize our calculation of the differential BGC rate $\frac{d\Gamma}{d\eion}(\eion)$ for collisions that impart $\eion$ kinetic energy onto the ion in the laboratory frame.  For details, see \appen~\ref{sec:collisionRate}.  To include motional heating due to glancing collision events, we calculate the differential cross section semiclassically following the approach of~\citet{Zipkes2011}.  Integrating the lab-frame differential cross section $\dsdW(\vartheta,v)$ with a 3D Maxwell-Boltzmann distribution of BG molecules $P(\vec{v})$ with temperature $T$ and number density $n$, we write the total scattering rate as
\begin{equation}\label{eq:scatteringRate}
	\Gamma = \int d^3 \vec{v} \ n v P(\vec{v})
		\int 2 \pi \sin \vartheta d\vartheta
		\frac{d \sigma}{d \Omega}\left(\vartheta,v\right) \ ,
\end{equation}
where $\vartheta$ and $v$ are the lab-frame scattering angle and initial BG velocity.  Finally, we substitute $\vartheta(\eion, v)$ based on elastic collision kinematics into Eq.~\ref{eq:scatteringRate} and numerically compute $\frac{d\Gamma}{d\eion}(\eion)$ via finite differences.

\begin{figure}
\begin{center}
\includegraphics[width=1.0\columnwidth]{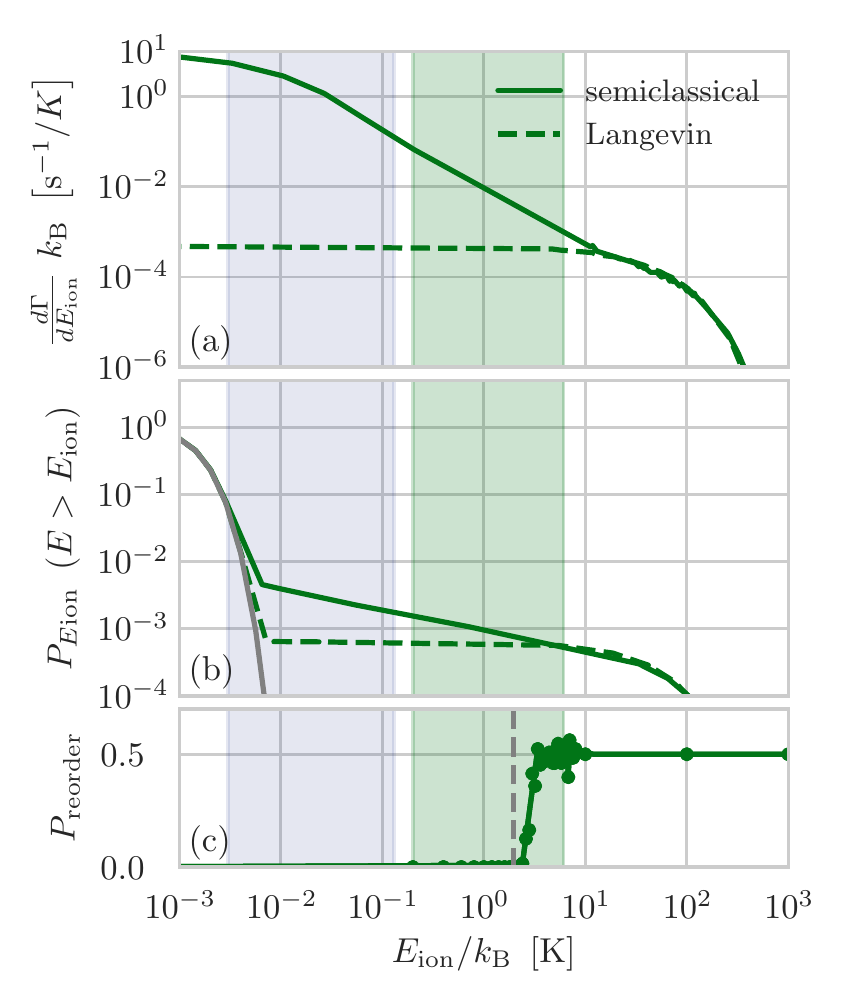}
\vspace{-0.7cm}
\caption{\label{fig:collisionRate}
	(Color online)
	(a)
	Differential rate of collisions between $^{27}$Al$^+$ at rest and 51.9~nPa of H$_2$ BG at 295~K resulting in final ion kinetic energy $\eion$. The cross section for Langevin spiraling collisions underestimates the collision rate for low energies when compared with the semiclassical calculation, which includes both glancing and spiraling collisions.
	(b)
	Cumulative energy distribution ($P_{\Eion}$) for all $E > \eion$.  We compare the Maxwell-Boltzmann distribution for a three dimensional harmonic oscillator at a typical Doppler cooling limit of 0.5~mK~(gray) with the distribution after 150~ms of collisions.
	(c)
	Example of a Monte-Carlo calculation of the $\caal$ reorder probability ($\preorder$) after $\alp$ receives $\eion$ energy due to a BGC. The potential energy barrier for reordering, $\ereorder$, is shown as a gray dashed line.  The green shaded region marks the experimentally accessible range of $\ereorder$ in this work.
	The blue shaded region indicates the range of $\rbybp$ collision energies explored by \citet{Zipkes2011}, mapped onto the resulting ion energy for H$_2$ -- $^{27}$Al$^+$ collisions.
}
\end{center}
\end{figure}

The collision-rate distribution predicted by our model for $\alp$ in 51.9~nPa of room temperature H$_2$ is shown in \autoref{fig:collisionRate}~(a).
The total collision rate of $0.16$~s${^{-1}}$ can be contrasted with a 0.020~s$^{-1}$
total collision rate predicted by the Langevin cross section~\cite{Wineland1998}. The effect of BGCs on the clock ion's kinetic energy after 150~ms of collisional heating is shown in \autoref{fig:collisionRate}~(b), indicating a highly nonthermal distribution for $\eion$ with $\approx 1~\%$ of the population distributed between $10^{-2}$ and $10^{2}$~K (energies throughout are are given in temperature units).
We calculate the collisional heating rate to be of order 1~K/s, which is significantly higher than the total heating rate inferred
by sideband thermometry measurements for the ion traps considered
here~\cite{Chen2017}. This discrepancy arises since typical sideband
thermometry assumes a thermal distribution and is
more sensitive to population near the motional ground state~\cite{Monroe1995}.  We have calculated the heating rate inferred by sideband thermometry for collisional heating to be of order 100~$\mu$K/s, or 0.1~quanta/s for each motional mode, for typical experimental parameters.

%%%%%%%%%%%%%%%%%%%%%%%%%%%%%%%%%%%%%%%%%%%%%%%%%%
% Two-ion mixed-species crystal collisional reordering rate
%%%%%%%%%%%%%%%%%%%%%%%%%%%%%%%%%%%%%%%%%%%%%%%%%%

For experiments using two-ion mixed-species crystals, it is possible to
measure the rate of collisions that cause the order of the crystal to change.
This measurement can be used to infer the pressure of the BG at the
position of the trapped ions, which can be significantly different from the
pressure in other locations of the UHV chamber, such as the
position of a pressure gauge or vacuum pump.
We calculate the reorder rate $\greorder$ by integrating the collision rate
with the reorder probability and summing over the two ions:
\begin{equation}\label{eq:reorderrate}
	\Gamma_\textrm{reorder} = \sum_{i = 1}^2 \int\limits_0^\infty \pireorder(\eion) \frac{d \Gamma_i}{dE_\textrm{ion}}(\eion) d E_\textrm{ion} \ ,
\end{equation}
where $\pireorder(\eion)$ is the reorder probability and $\frac{d \Gamma_i}{d
E_\textrm{ion}}(\eion)$ is the differential collision rate for ion $i$.
While an accurate evaluation of $\pireorder(\eion)$ requires detailed knowledge of
the ion trajectories following a collision, a simpler expression can be
obtained with the assumption that $\pireorder(\eion) = 0.5$ for $\eion > \ereorder$.  Here, $\ereorder$ is the potential-energy barrier between a ground-state axial
configuration and a two-ion radial crystal, which depends on the trapping conditions (\appen~\ref{sec:reorderBarrier}). This reduces
\autoref{eq:reorderrate} to an integral over all collision energies that are
greater than $\ereorder$ similar to the cumulative distribution shown in
\autoref{fig:collisionRate}~(b) and approximately given by
\begin{equation}\label{eq:reorderrateApprox}
\Gamma_\textrm{reorder} \lessapprox \frac{1}{2} \left( \frac{p}{902~\textrm{nPa$\,\cdot\,$s} } \right) \left( \frac{\ereorder}{\mathrm{1~K\times\kb}} \right)^{-0.278}
\end{equation}
for $\caalp$ in 295~K
H$_2$ background gas and 0.1~K$< \ereorder <$10~K, where $p$ is the H$_2$
pressure. The constants given in \autoref{eq:reorderrateApprox} are specific to
the background-gas temperature as well as the masses of the two ions.

\begin{figure}
	\begin{center}
	\includegraphics{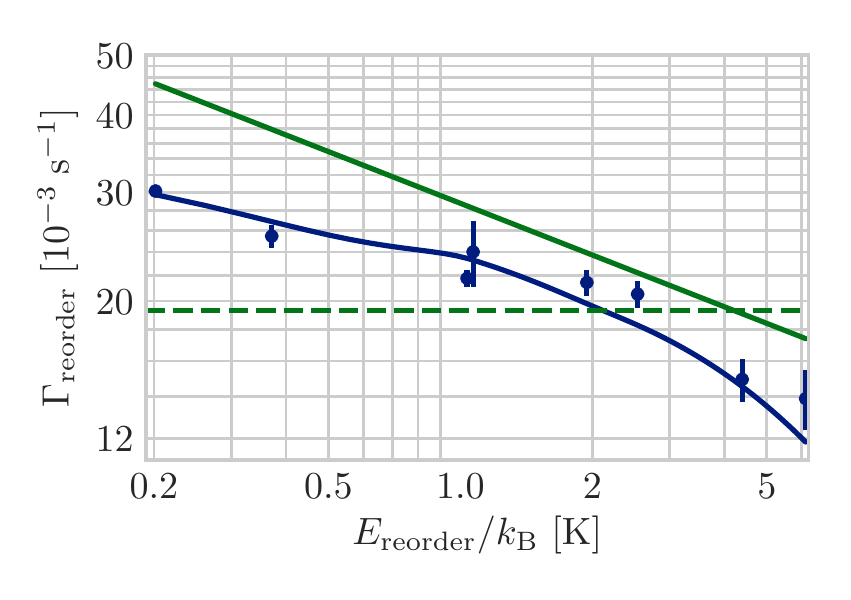}
	\caption{\label{fig:reorderRate}
		(Color online)
		Reorder rate as a function of reorder energy barrier for a $\caalp$ two-ion
		crystal in H$_2$ background gas at 295~K.
		Blue points:
		experimental data.
		Blue line:
		single-parameter fit calculated using \autoref{eq:reorderrate} together
		with trajectory simulations for $P_{i, \mathrm{reorder}}(\eion)$.  The estimated
		background gas pressure is $51.9$~nPa, with a $1 \sigma$ statistical uncertainty of
		$\pm 0.8$~nPa and a $\chi_{\mathrm{red}}^2$ of $2.4$
		(see \appen~\ref{sec:CollisionShift} for a discussion of the systematic uncertainty).
		Green line:
		predicted upper bound on the reorder rate (\autoref{eq:reorderrateApprox}) at a pressure of $51.9$~nPa. 
		Dashed green line:
		reorder rate for the same pressure when assuming 50\% of Langevin collisions cause the crystal to reorder, as was used in previous clock evaluations. 
		}
	\end{center}
\end{figure}

%%%%%%%%%%%%%%%%%%%%%%%%%%%%%%%%%%%%%%%%%%%%%%%%%%
% Experimental validation of the differential cross section
%%%%%%%%%%%%%%%%%%%%%%%%%%%%%%%%%%%%%%%%%%%%%%%%%%

In order to verify the nonthermal energy distribution predicted by our model
for $\frac{d \Gamma_i}{d E_\textrm{ion}}(\eion)$, we measure two-ion reorder rates for experimentally accessible
values of $\ereorder$ in a $\caalp$ clock. The setup is similar to the $\mgalp$ clock described in Refs.~\cite{Chen2017,Chen2017thesis,Brewer2019} with the sympathetic cooling and quantum-logic readout ion $^{25}$Mg$^+$ replaced by $\cap$.  Two-ion reorder events are detected by monitoring the position of the $\cap$ ion with respect to the $\alp$ ion.
The logic ion is detected and cooled with a 397-nm laser tuned near resonance with the
$|S_{1/2}\rangle\rightarrow |P_{1/2}\rangle$ cycling transition.  Using a
matched pair of aspherical lenses with a numerical aperture of 0.38, $\ca$
fluorescence is collected and relayed onto a knife-edge spatial filter aligned
to the center of the image of the two-ion crystal. The spatially filtered
image is then relayed onto a photomultiplier tube using a microscope
objective, resulting in two discrete fluorescence levels for the two 
crystal orders (\appen~\ref{sub:orderdetection}).  The reorder barrier is adjusted by
changing the dc voltage applied to the rf-Paul trap
endcaps~(\appen~\ref{sec:reorderBarrier}).  

The measured reorder rates for $\ereorder$ ranging from 0.2 to 6~K are shown in
\autoref{fig:reorderRate}. While \autoref{eq:reorderrateApprox} 
predicts the overall trend, a more accurate estimate of $\greorder$ can be
calculated by considering the operating parameters of the ion trap and the
details of the laser cooling after a collision.  To include these effects, we
have calculated $\pireorder(\eion)$ for the experimental values of $\ereorder$ using
Monte-Carlo simulations of the ion trajectories after a
collision in the full time-dependent potential~(\autoref{fig:collisionRate}~(c), \appen~\ref{sec:trajectorSims}).  A single-parameter fit of this model
for pressure gives 51.9(8)~nPa, where the error bar indicates the $1 \sigma$ statistical uncertainty, with $\chi_{\mathrm{red}}^2 = 2.4$.  In the following, we include a 50~\% systematic uncertainty in addition to the statistical error in our estimate of the pressure uncertainty, to account for inelastic collision processes and the unknown short-range interaction potential (see \appen~\ref{sec:CollisionShift}).  The measured ion pump current indicated a pressure of 20(10)~nPa when the UHV chamber was first built and 100(20)~nPa during the reorder measurements.
Fitting \autoref{eq:reorderrateApprox} to the data results in a pressure of $38.9(6)$~nPa with $\chi_{\mathrm{red}}^2 = 10$, indicating that ion trajectory calculations are necessary for accurate pressure determination. Both approaches improve upon the model used in past $\alp$ clock uncertainty evaluations that did not consider glancing collisions and assumed all Langevin collisions cause the ions to reorder with 50~\% probability~\cite{Rosenband2008} (\autoref{fig:reorderRate} dashed green line).  If this model is fit to the reorder data, it demonstrates poor agreement with a $\chi_{\mathrm{red}}^2$ of 31.

Our measurements are in agreement with the collision kinematics model described above over the experimentally accessible energy range $0.2~\mathrm{K} < \eion < 6~\mathrm{K}$ indicated by the green shaded region in \autoref{fig:collisionRate}, which roughly corresponds to a verification of our semiclassically calculated differential cross section over a range of lab-frame scattering angles between 0.09 and 0.5~rad, for collisions between $\alp$ at rest and room-temperature H$_2$ gas.  \citet{Zipkes2011} have studied the collision kinematics of a single $^{172}$Yb$^+$ ion immersed in a cloud of ultracold $^{87}$Rb atoms.  They find that Langevin collisions alone are insufficient to explain the observed collisional heating and loss of the neutral atoms, and that it is necessary to calculate the differential cross section semiclassically including both Langevin and glancing collisions.  By measuring the collisional heating and loss of $^{87}$Rb as a function of the initial kinetic energy of $^{172}$Yb$^+$, they verify the same collision kinematics model that we use, but over a different energy range indicated by the blue shaded region in \autoref{fig:collisionRate}.  Together, these two experiments verify the collision kinematics model over nearly the entire range of resulting ion energies that contribute to the BGC shift of trapped-ion clocks.

%%%%%%%%%%%%%%%%%%%%%%%%%%%%%%%%%%%%%%%%%%%%%%%%%%
% Optical atomic clock BGC shift
%%%%%%%%%%%%%%%%%%%%%%%%%%%%%%%%%%%%%%%%%%%%%%%%%%

An upper bound on the time-dilation component of the BGC shift can be obtained based on the collisional heating rate discussed above.  For the $^{27}$Al$^+$ clocks considered here, operated without sympathetic cooling during a 150~ms probe, the time-averaged fractional time-dilation shift is of order $10^{-16}$ .  It is important to recognize, however, that this is dominated by a very low population tail of the motional state distribution at very high energy.  In practice, the contribution of population at very high energy to the spectroscopy signal is suppressed both because the transition is shifted completely off-resonance by the time-dilation effect, and because the Rabi frequency is suppressed by motional Debye-Waller factors \cite{Wineland1998,Chen2017}.

To take this suppression into account, we calculate the BGC shift using Monte-Carlo simulations of the spectroscopy sequence including the ion motion.  We assume that for each probe the two ions begin in the three-dimensional motional ground state, and there is no additional cooling or heating during the spectroscopy pulse \cite{Chen2017}.  Note that this is not applicable to previous $\alp$ clocks which operated with continuous sympathetic cooling during the spectroscopy pulse \cite{Rosenband2008,Chou2010}.  Following \citet{DeVoe2009}, for each of the typically $10^8$ Monte-Carlo trials, we analytically propagate the classical motion of the ions between collisions to first-order in the Mathieu $q$ parameter.  For each ion, we randomly generate collision times from an exponential distribution with a mean given by the reciprocal of the total elastic collision rate $\Gamma$.  For each collision, we randomly choose $\eion$ from a distribution obtained by normalizing $\frac{d \Gamma_i}{d \eion}(\eion)$.  From the amplitudes and phases of the motional modes before the collision we calculate the instantaneous velocity of the colliding ion, add $\Delta v = \sqrt{2 \eion/m_\textrm{ion}}$ with a randomly selected direction, and calculate the amplitudes and phases of the motional modes after the collision.  Here, $m_\textrm{ion}$ is the mass of the colliding ion.  For each time during the probe, the classical oscillation amplitudes are mapped onto a 6D coherent state, and Fock states are randomly selected from the corresponding distributions.  Using these time-dependent Fock states for each motional mode, we analytically propagate the internal quantum superposition of the clock ion under a Rabi spectroscopy Hamiltonian for a grid of values of the laser frequency covering the atomic transition, including the second-order time dilation shift of the atomic transition frequency and the Debye-Waller suppression of the Rabi frequency.  

\begin{table}
\caption{\label{tab:collisionShift} Frequency shift and uncertainty of the $\mgalp$ optical atomic clock presented in \cite{Brewer2019} due to background-gas collisions.  In the column labels, ``suppression'' refers to the combined effect of Debye-Waller reduction of Rabi frequencies and the shift of the transition frequency of high energy population outside of the Fourier-limited lineshape. Past $\alp$ optical clock evaluations have only considered the phase shift part of the BGC shift without suppression~\cite{Rosenband2008}.
}
\begin{ruledtabular}
\begin{tabular}{lll}
Contribution					& \multicolumn{2}{l}{Fractional frequency shift [$10^{-19}$]} \\
							& Without suppression			& With suppression \\
\hline
Time-dilation\rule{0pt}{3ex}	& $-2200 \pm 1100$				& $-0.6^{+0.6}_{-0.3}$ \\
Phase shift					& \phantom{$-220$}$0 \pm 42$	& \phantom{$-.6$}$0 \pm 2.3$ \\
Total						& $-2200 \pm 1100$				& $-0.6 \pm 2.4$ \\
\end{tabular}
\end{ruledtabular}
\end{table}

For each collision involving the clock ion, we randomly decide if the event is a Langevin spiraling collision based on the ratio of the differential Langevin collision rate (\appen~\ref{sec:differentialCrossSection}) to $\frac{d \Gamma_i}{d \eion}(\eion)$, and for Langevin spiraling collisions that penetrate the angular momentum barrier we add a worst case $\pm \pi/2$ phase shift to the atomic superposition state \cite{Rosenband2008} with an unknown sign that is fixed for each set of trials.
In general, there is also a phase shift for glancing collisions because the ion polarizes the molecule, and the resulting dipole electric field Stark shifts the ion.  However, for $\alp$, because of the small differential polarizibility of the clock transition~\cite{Safronova2011,Brewer2019}, the frequency shift caused by glancing collision phase shifts is negligible compared with the frequency shift caused by Langevin collision phase shifts.
At the end of the Rabi probe time, we calculate the transition lineshape as the population in the excited state for each laser frequency, and we average this lineshape over all of the Monte-Carlo trials.  Finally, we calculate the fractional BGC shift of the clock, $\nu_c/\nu$, by finding the center frequency of the lineshape $\nu_c$ defined as the average of the two frequencies where the transition probability is equal to half of the maximum transition probability.

For the $\mgalp$ quantum-logic clock presented in \cite{Brewer2019}, with 38(19)~nPa of H$_2$ BG (measured in situ as described above for the $\caalp$ clock) at $294.15(2.70)$~K, for a 150~ms probe duration the Monte-Carlo simulations constrain the collisional frequency shift to be $-0.6(2.4)\times 10^{-19}$ (\appen~\ref{sec:CollisionShift}).  Table~\ref{tab:collisionShift} shows the time-dilation and the collisional phase shift components separately.  As noted earlier, without the suppression of the contribution of high energy population to the spectroscopy signal, the collisional heating time-dilation shift is of order $10^{-16}$, but this is shown to be suppressed by more than three orders of magnitude by the Monte-Carlo simulations.  The atomic phase shift associated with the worst case $\pm \pi/2$ bound is smaller but features less suppression due to forward scattering Langevin collisions, and dominates the final uncertainty. The $2.4\times 10^{-19}$ uncertainty of the BGC shift calculation described here is substantially smaller than would be obtained by applying the analysis method previously used by~\citet{Rosenband2008}, which yields an uncertainty of $4\times 10^{-18}$ for this clock. Inclusion of calculated potential energy surfaces for an AlH$_2^+$ molecule in our model would further reduce the BGC uncertainty without the need for an improvement to the vacuum pressure \cite{Vutha2017}, although the effect of small amounts of BG species other than H$_2$ would have to be considered.  

\begin{figure}[h]
	\begin{center}
		\includegraphics[width=1.0\columnwidth]{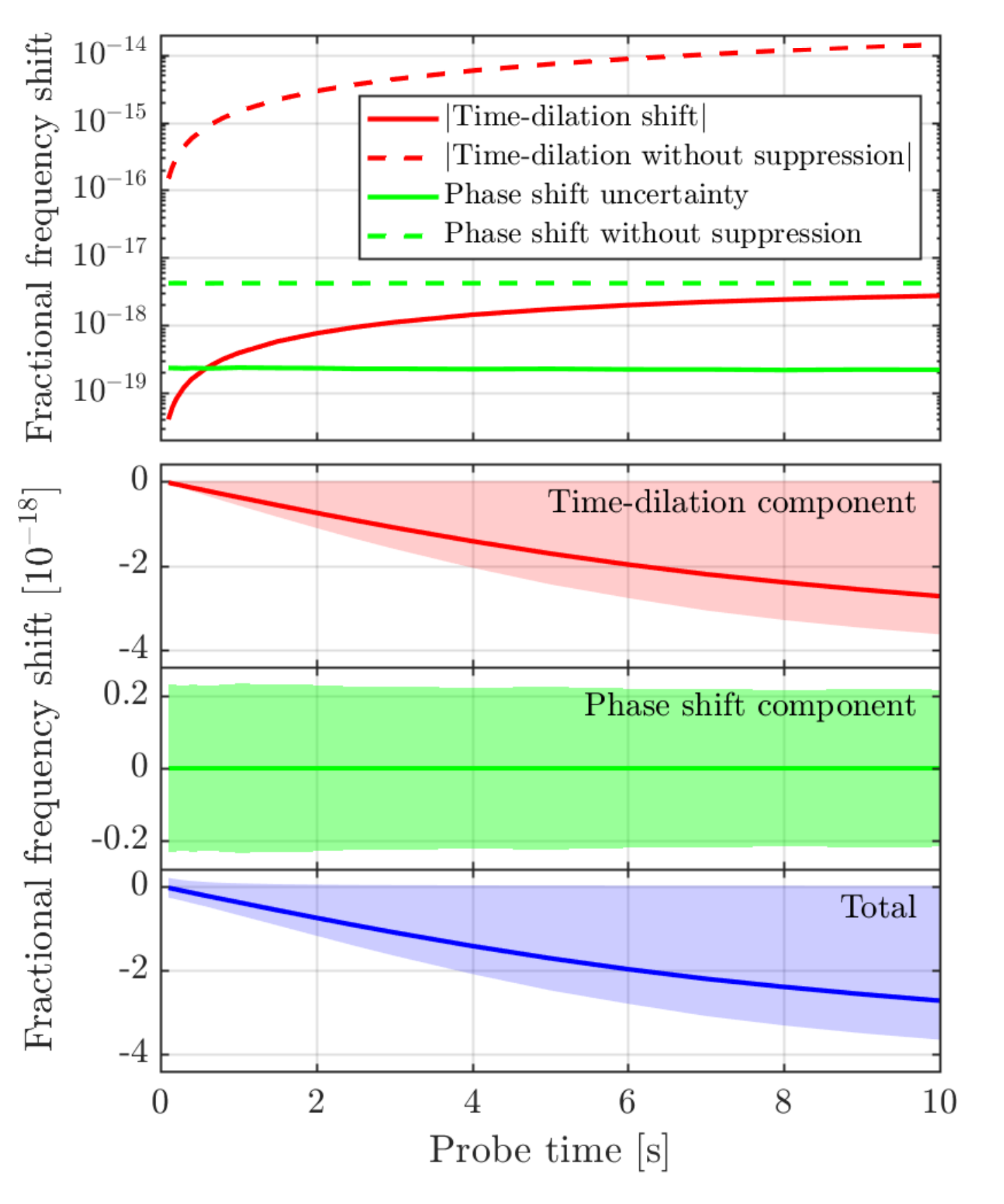}
		\caption{\label{fig:collisionShiftVsProbeTime}(Color online) Fractional BGC shift and uncertainty of the $^{27}$Al$^+$ clock presented in \cite{Brewer2019} as a function of the spectroscopy probe time.  (top panel) Time-dilation component of the frequency shift (red) and uncertainty due to the phase shift component (green) with (solid lines) and without (dashed lines) including suppression due to the combined effect of Debye-Waller reduction of Rabi frequencies and the time-dilation shift of the transition center frequency outside of the Fourier-limited lineshape in the Monte-Carlo model.  (bottom panels) The red, green, and blue lines and shaded regions show the shift and uncertainty of the time-dilation component, the phase shift component, and the total BGC shift.
		}
	\end{center}
\end{figure}

The time-dilation component of the BGC shift will grow with the probe duration as is shown in Fig.~\ref{fig:collisionShiftVsProbeTime}.  The time-dilation component scales linearly with the clock interrogation time when this probe time is much less than the collision period, then begins to flatten out as it becomes more likely to have multiple collisions in a single probe. The phase shift component is approximately constant due to a balance of the increasing probability of having a Langevin spiraling collision during each probe and the decreasing spectroscopy linewidth (and hence maximum frequency shift due to a phase shift) as the probe time increases.  For seconds-long probe times, which are attainable using more stable clock lasers \cite{Matei2017} or correlation spectroscopy techniques \cite{Chou2011,Hume2016}, the calculated collision shift would dominate the systematic uncertainty of the clock described in \cite{Brewer2019} under these conditions.

\begin{table*}
	\caption{\label{tab:clockParameters} Operating parameters and calculated collisional frequency shifts for $^{27}$Al$^+$, $^{40}$Ca$^+$, $^{88}$Sr$^+$, and $^{171}$Yb$^+$ clocks.  For $^{27}$Al$^+$, we use the parameters of and calculate the BGC shift and uncertainty specifically for the clock described in \cite{Brewer2019}.  For the $^{40}$Ca$^+$, $^{88}$Sr$^+$, and $^{171}$Yb$^+$ clocks, we use parameters similar to experimental implementations at WIPM \cite{Huang2016,Huang2017}, NRC \cite{Dube2016,Dube2013}, and PTB \cite{Huntemann2016}, but we do not perform a rigorous analysis of the uncertainty of the BGC shift.  For all clocks, we simulate Rabi interrogation.  Current $^{171}$Yb$^+$ clocks based on the 467~nm electric-octupole (E3) transition use hyper-Ramsey \cite{Huntemann2016} or autobalanced Ramsey \cite{Sanner2018} interrogation, so this calculated collision shift for $^{171}$Yb$^+$ should be taken as a rough estimate.}
	\begin{ruledtabular}
		\begin{tabular}{lllll}
			& $^{27}$Al$^+$
			& $^{40}$Ca$^+$
			& $^{88}$Sr$^+$
			& $^{171}$Yb$^+$ E3 \\
			
			\hline
			Trap drive frequency [MHz]
			& \phantom{0}40.7
			& \phantom{0}24.7
			& \phantom{0}14.4
			& \phantom{00}15 \\
			
			Secular frequencies [MHz], x-direction
			& \phantom{00}2.85, 3.31
			& \phantom{00}1.01
			& \phantom{00}1.13
			& \phantom{000}0.7 \\
			
			\phantom{Secular frequencies [MHz],} y-direction
			& \phantom{00}3.54, 3.95
			& \phantom{00}1.07
			& \phantom{00}1.15
			& \phantom{000}0.7 \\
			
			\phantom{Secular frequencies [MHz],} z-direction (axial)
			& \phantom{00}1.47, 2.55
			& \phantom{00}2.31
			& \phantom{00}2.25
			& \phantom{000}1.4 \\
			
			Interrogation time [ms]
			& \phantom{}150
			& \phantom{0}80
			& \phantom{}100
			& \phantom{0}150
			\\
			
			Clock laser direction, $\hat{k}$
			& \phantom{}$\left( 1/2, 1/2, 1/\sqrt{2} \right)$
			& \phantom{}$\left( 1, 0, 1 \right)/\sqrt{2}$
			& \phantom{}$\left( 1, 1, 0 \right)/\sqrt{2}$
			& \phantom{0}$\left( 1, 1, 0 \right)/\sqrt{2}$
			\\
			
			BG temperature [K]
			& \phantom{}$294.15 \pm 2.70$
			&	\phantom{}300
			&	\phantom{}300
			&	\phantom{0}300
			\\
			
			BG pressure [nPa]
			& \phantom{0}$38 \pm 19$
			& \phantom{0}30
			& \phantom{0}16
			& \phantom{00}6
			\\
			
			\hline
			Fractional frequency shift, time-dilation [$10^{-19}$] \rule{0pt}{3ex}
			& \phantom{0}$-0.6^{+0.6}_{-0.3}$
			& \phantom{0}$-0.9$
			& \phantom{00}-0.2
			& \phantom{-00}0.0
			\\
			
			\phantom{Fractional frequency shift,} phase shift [$10^{-19}$]
			& \phantom{0}$\pm 2.3$
			& \phantom{0}$\pm 13.3$
			& \phantom{0}$\pm 11.9$
			& \phantom{0}$\pm 2.5$
			\\
			
			\phantom{Fractional frequency shift,} total [$10^{-19}$]
			& \phantom{0}$-0.6 \pm 2.4$
			& \phantom{0}$-0.9 \pm 13.3$
			& \phantom{0}$-0.2 \pm 11.9$
			& \phantom{-00}$0.0 \pm 2.5$
			\\
		\end{tabular}
	\end{ruledtabular}
\end{table*}

%%%%%%%%%%%%%%%%%%%%%%%%%%%%%%%%%%%%%%%%%%%%%%%%%%
% Conclusion
%%%%%%%%%%%%%%%%%%%%%%%%%%%%%%%%%%%%%%%%%%%%%%%%%%

In summary, we have developed a framework for calculating the BGC shift for trapped-ion optical clocks that includes both the time-dilation shift due to the kinetic energy received during a BGC as well as the phase shift of the atomic superposition state. Using the semiclassical differential scattering cross section, we show that the inclusion of glancing collisions is important for accurately describing the frequency of collisions that deposit kinetic energies of $< 10$~K to the ions and therefore necessary for accurately evaluating the clock ion's kinetic energy.
We have verified the predicted collisional energy distribution by measuring the rate of two-ion reorder events in a $\caalp$ experiment due to BGCs over the experimentally accessible range of reorder barriers ranging from 0.2 to 6~K. Finally, we performed a Monte-Carlo simulation of spectroscopy of the 267~nm $\alp$ $\sztopz$ clock transition for the $\mgalp$ quantum-logic clock presented in \cite{Brewer2019} to calculate the BGC shift, $-0.6(2.4)\times 10^{-19}$, enabling total clock uncertainty below $10^{-18}$ despite a BG pressure that is a factor of two to three higher than previous $\alp$ clocks.

The suppression of the BGC shift and uncertainty due to Debye-Waller factors
will be smaller for clocks with visible or infrared optical clock transitions
due to their smaller Lamb-Dicke parameters, and is negligible in trapped-ion
microwave clocks.  We have estimated the BGC uncertainty using our framework to be
near $10^{-18}$ for $^{40}$Ca$^+$, $^{88}$Sr$^+$, and
$^{171}$Yb$^+$ clocks with typical experimental parameters
(\autoref{tab:clockParameters}).  We thus expect the BGC
uncertainty modeled here to be significant in most room-temperature trapped-ion
clocks as their systematic uncertainties reach the $10^{-18}$ level or
below.

The BGC heating of trapped ions considered here and the pressure measurement technique we use (similar to that described by \citet{Pagano2019}) may be relevant to many precision measurement, quantum information \cite{Wineland1998,Chiaverini2014,Pagano2019,Jurcevic2017}, and ultracold atom-ion mixture \cite{Zipkes2011,Ratschbacher2013,Meir2016,Schowalter2016} 
experiments.  By incorporating fully quantum scattering calculations, it may be possible to use this pressure measurement technique as the basis of a primary pressure standard.  Due to the months long trapping times possible with trapped ions, such a standard may have a significantly lower measurement floor than those based on neutral atoms \cite{Scherschligt2017}.

We thank T.~Rosenband (Harvard University) for development of the ion trajectory integrator, A.~Vutha (University of Toronto) for useful discussions, and K.~Beloy and E.~Tiesinga (National Institute of Standards and Technology) for their careful reading of the manuscript.  This work was supported by the Defense Advanced Research Projects Agency, the National Institute of Standards and Technology, and the Office of Naval Research.  S.M.B. was supported by the U.S. Army Research Office through MURI Grant No.~W911NF-11-1-0400.  This Letter is a contribution of the U.S. Government, not subject to U.S. copyright.

\textit{Note added} --- During preparation of this manuscript, we became aware of the work by \citet{Davis2019} in which the state dependent potential energy surfaces and partial wave scatting phase shifts for Al$^+$ -- H$_2$ collisions are calculated.

%%%%%%%%%%%%%%%%%%%%%%%%%%%%%%%%%%%%%%%%%%%%%%%%%%
\appendix
\section{Background-gas collisions}\label{sec:collisionRate}
%%%%%%%%%%%%%%%%%%%%%%%%%%%%%%%%%%%%%%%%%%%%%%%%%%

In the following subsections, we calculate the scattering differential cross section using the semiclassical partial-wave expansion~\cite{Sakurai1994} and the resulting kinetic-energy distribution of the ion including both glancing and spiraling collisions.

%%%%%%%%%%%%%%%%%%%%%%%%%%%%%%%%%%%%%%%%%%%%%%%%%%
\subsection{Differential cross section}\label{sec:differentialCrossSection}
%%%%%%%%%%%%%%%%%%%%%%%%%%%%%%%%%%%%%%%%%%%%%%%%%%

At large separations $R$ the interaction potential between a neutral BG
molecule and an ion is given by 
\begin{equation}
V(R) = -\frac{C_4}{R^4} \ ,
\end{equation}
where $C_4 = \alpha Q^2 / (8 \pi \epsilon_0)$, $\alpha$ is the polarizability
of the BG molecule, $Q$ is the charge of the ion, and $\epsilon_0$
is the vacuum permittivity \cite{Wineland1998}. For H$_2$, the orientation-averaged static polarizability $\alpha = 0.8 \times 10^{-30}$~m$^3$
\cite{Landolt1951,Akindinova2009,Dalgarno1958}.  At small separations the interaction
potential becomes repulsive and state dependent, but the exact form depends on the collision
species and is not well known for the species considered here.  Impact
parameters smaller than the Langevin radius $b_\textrm{crit} = (\alpha Q^2 /
\pi \epsilon_0 \mu v^2)^{1/4}$ result in Langevin collision trajectories in which the molecule spirals in towards the ion.  Here, $\mu =
m_\textrm{bg} m_\textrm{ion} / (m_\textrm{bg} + m_\textrm{ion})$ is the
reduced mass, $m_\textrm{bg}$ is the mass of the BG molecule,
$m_\textrm{ion}$ is the mass of the ion, and $v$ is the relative velocity.  For
H$_2$ at a temperature $\mathrm{T} = 300$~K, the critical impact parameter for
the mean speed $\left<v\right> = \sqrt{\frac{8 k_B T}{\pi m_\textrm{bg}}} =
1.8$~km/s where $k_B$ is the Boltzmann constant is 0.5~nm.  

Following \citet{Zipkes2011}, the differential cross section in the center-of-mass coordinate system is calculated semiclassically as a sum over partial waves,
\begin{equation}\label{eq:partialwavesum}
\frac{d\sigma'}{d\Omega'} (\theta) = \frac{1}{k^2} \left| \sum_{l=0}^{\infty} (2 l + 1) e^{i \eta_l} \sin(\eta_l) P_l(\cos \theta) \right|^2 \ ,
\end{equation}
where the angular and linear momentum of a partial wave are $\hbar l$ and
$\hbar k = \sqrt{2 \mu E_c}$, $E_c = \mu v^2/2$ is the collision energy,
$\eta_l$ is the scattering phase, $P_l(\cos \theta)$ is the $l$th order
Legendre polynomial, and $\theta$ is the scattering angle in the center-of-mass
inertial frame.  Partial waves with $l < l_\textrm{crit} = \sqrt{2 \mu \sqrt{4
C_4 E_c}}$ have kinetic energy larger than the centrifugal potential barrier and sample the repulsive
short range potential, resulting in rapidly oscillating scattering phases \cite{Cote2000}.  
Without detailed knowledge of the short range potential for the collision partners considered here, we
assume that $\eta_l$ are uniformly distributed random numbers over the range 0
to $2 \pi$ for $l < l_\textrm{crit}$, and we average the resulting differential
cross section over many instances of the random scattering phases.
For $l > l_\mathrm{crit}$, we use the semiclassical
approximation for the phase shift
\begin{equation}
	\eta_l = - \frac{\mu}{\hbar^2} \int_{R_\mathrm{crit}}^{\infty} \frac{V(R)}{\sqrt{k^2 - (l + 1/2)^2/R^2}} dR \approx \frac{\pi \mu^2 C_4 E_c}{2 \hbar^4 l^3} \ ,
\end{equation}
where $R_\mathrm{crit} = (l + 1/2)/k$ is the classical turning point.  
The resulting differential cross section is consistent with the alternative treatment of \citet{Dalgarno1958},
has been shown to
describe the kinematics of $^{87}$Rb - $^{172}$Yb$^+$ collisions
\cite{Zipkes2011}, and is further supported by the experiments presented in the main text.  

\begin{figure}
\begin{center}
	\includegraphics[width=1.0\columnwidth]{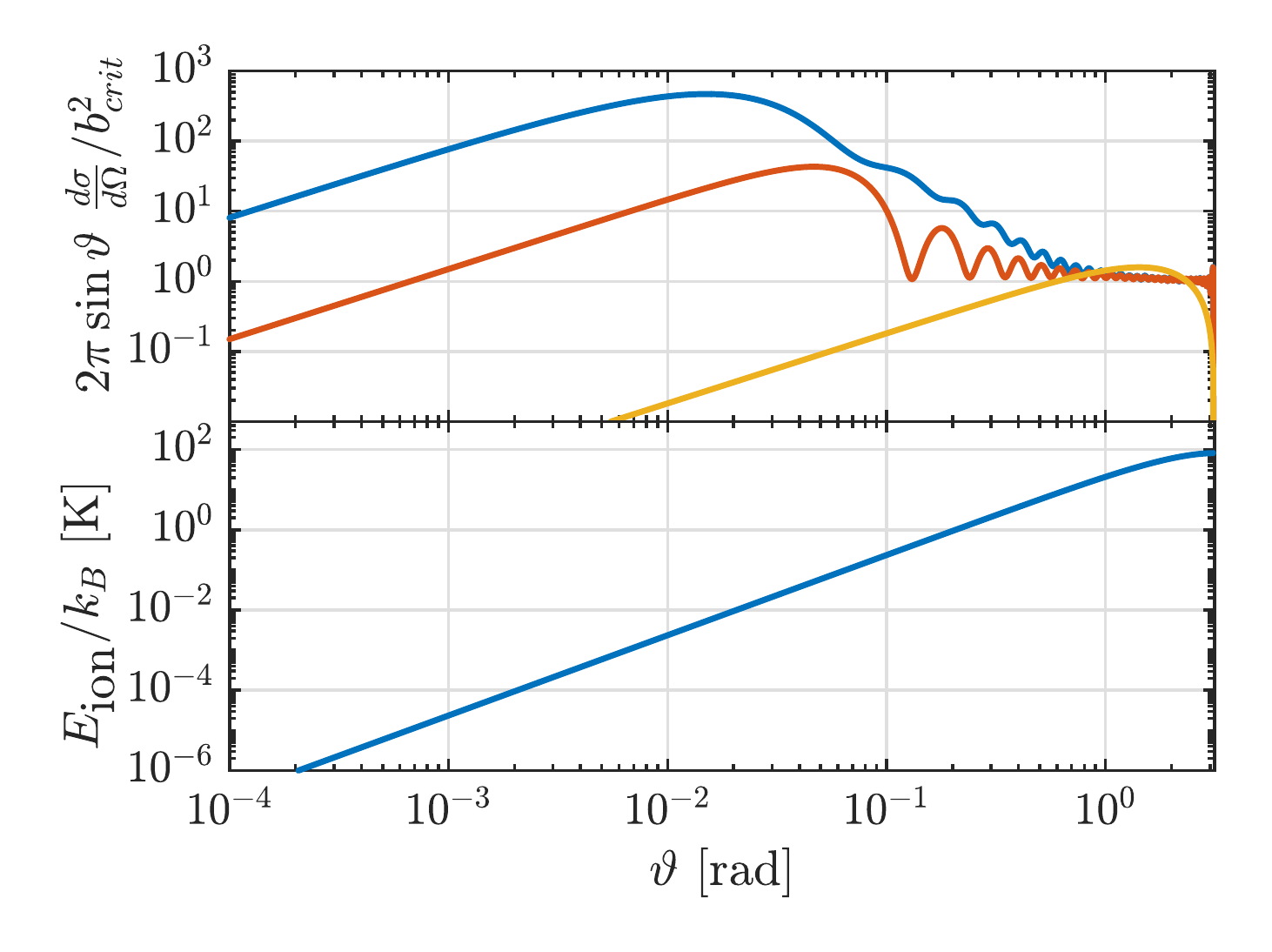}
\caption{\label{fig:differentialCrossSection}(Color online) (top panel) Lab-frame differential cross section for collisions between $^{27}$Al$^+$ at rest and H$_2$ with 300~K of kinetic energy as a function of scattering angle. 
Blue: semiclassical differential cross section including both glancing and spiraling collisions.  Red: semiclassical differential cross section including only spiraling collisions, calculated by setting $\eta_l = 0$ for $l > l_{\mathrm{crit}}$.
Yellow: classical Langevin differential cross section assuming uniform scattering in the center-of-mass frame and that the total cross section is given by $\pi b_{\mathrm{crit}}^2$.
(bottom panel) Lab-frame kinetic energy of the $^{27}$Al$^+$ ion after the collision as a function of scattering angle.}
\end{center}
\end{figure}

We transform from the center-of-mass frame differential cross section $\frac{d \sigma'}{d \Omega'}(\theta)$ to the laboratory frame differential cross section $\frac{d \sigma }{d \Omega}(\vartheta)$ using \cite{Goldstein2002}
\begin{equation}
	\vartheta = \arccos \left( \frac{\cos \theta + \rho}{\sqrt{1 + 2 \rho \cos \theta + \rho^2}} \right)
\end{equation}
and
\begin{equation}
\frac{d \sigma}{d \Omega} (\vartheta) = \frac{d \sigma'}{d \Omega'} (\theta) \frac{(1 + 2 \rho \cos \theta + \rho^2)^{3/2}}{1 + \rho \cos \theta}
\end{equation}
where $\rho = m_\textrm{bg}/m_\textrm{ion}$.  For the experiments considered here, we perform laser cooling on a timescale much faster than the total elastic collision rate, so the ion can be assumed to be at rest in the lab frame before the collision.  The kinetic energy of the ion in the lab frame after the collision is given by
\begin{equation}\label{eq:ionEnergy}
E_\textrm{ion} = \frac{m_\textrm{bg} v^2}{2} \frac{2 \rho}{(1 + \rho)^2} (1 - \cos \theta) \ .
\end{equation}
Fig.~\ref{fig:differentialCrossSection} shows the differential cross section and final ion energy for collisions between $^{27}$Al$^+$ at rest and H$_2$ with 300~K of kinetic energy.

For the BGC shift Monte-Carlo simulations described in the main text, we also require the differential cross section for Langevin spiraling collisions (i.e., not including glancing collisions).  We calculate this semiclassically using the method described above with $\eta_l$ set to zero for $l > l_{\mathrm{crit}}$, with the result shown in Fig.~\ref{fig:differentialCrossSection}.  For the small scattering angles which have the largest contribution to the BGC shift (Appendix~\ref{sec:BGCShift}), this reproduces the semiclassical differential cross section for a hard sphere with radius $b_\textrm{crit}$ \cite{Sakurai1994}.  Fig.~\ref{fig:differentialCrossSection} also shows the classical differential cross section for Langevin spiraling collisions under the assumption that they scatter uniformly into $4\pi$ with a total cross section of $\pi b_\textrm{crit}^2$ \cite{Chen2014}.  For small scattering angles, the classical differential cross section is more than two orders of magnitude smaller than the semiclassical one, indicating that our differential cross section for Langevin spiraling collisions is a more conservative estimation of the resulting phase shift component of the BGC shift.\\

%%%%%%%%%%%%%%%%%%%%%%%%%%%%%%%%%%%%%%%%%%%%%%%%%%
\subsection{Scattering rate distribution}\label{sec:scatterRate}
%%%%%%%%%%%%%%%%%%%%%%%%%%%%%%%%%%%%%%%%%%%%%%%%%%

We calculate the total elastic collision rate by integrating the differential cross section with a Maxwell Boltzmann distribution of BG velocities:
\begin{equation} \label{eq:totalScatRate}
\Gamma = \int_{\vec{v}} d^3 \vec{v} \ n \left( \frac{m_\textrm{bg}}{2 \pi k_B T} \right)^{3/2} e^{-\frac{m_\textrm{bg} v^2}{2 k_B T}} v \int_{\Omega} d\Omega \frac{d \sigma}{d \Omega} (\vartheta, v) \ ,
\end{equation}
where $n$ and $T$ are the number density and temperature of the BG.
For $\alp$ in 51.9~nPa of room temperature H$_2$, the total elastic collision rate is 0.16~s$^{-1}$.  This can be
contrasted with the classical rate of spiraling Langevin collisions $n \pi
b_\textrm{crit}^2 v = 0.020$~s$^{-1}$. The differential BGC rate $\frac{d\Gamma}{d\eion}(\eion)$ for collisions that impart $\eion$ kinetic energy onto the ion in the laboratory frame is given by
\begin{widetext}
	\begin{equation}\label{eq:diffCollisionRate}
	\dgde(\eion) = \int_{\vec{v}} d^3 \vec{v} \ n \left( \frac{m_\textrm{bg}}{2 \pi k_B T} \right)^{3/2} e^{-\frac{m_\textrm{bg} v^2}{2 k_B T}} v \frac{2 \pi \sin \left( \vartheta(\eion, v) \right)}{\left. \frac{d E_\textrm{ion}}{d \vartheta} \right|_{\vartheta(\eion, v)}} \frac{d \sigma}{d \Omega} (\vartheta(\eion, v), v) \ ,
\end{equation}
\end{widetext}
where $\frac{d \eion}{d \vartheta}$ and $\vartheta(\eion, v)$ are the derivative and the inverse of
Eq.~\ref{eq:ionEnergy} when written as a function of the lab frame scattering
angle $\vartheta$. In practice, we calculate $\frac{d\Gamma}{d\eion}(\eion)$ numerically as the value of Eq.~\ref{eq:totalScatRate} with the range of the inner integral restricted to $[\vartheta(\eion-d\eion/2, v), \vartheta(\eion+d\eion/2, v)]$ divided by $d\eion$.

%%%%%%%%%%%%%%%%%%%%%%%%%%%%%%%%%%%%%%%%%%%%%%%%%%
\section{Two-ion mixed-species crystal collisional reordering rate}\label{sec:twoIonCyrstal}
%%%%%%%%%%%%%%%%%%%%%%%%%%%%%%%%%%%%%%%%%%%%%%%%%%

In the following, we discuss the kinematics of a two-ion crystal undergoing a collision event and describe Monte-Carlo trajectory simulations used to calculate the two-ion reorder probability, $\preorder(\eion)$.  

%%%%%%%%%%%%%%%%%%%%%%%%%%%%%%%%%%%%%%%%%%%%%%%%%%
\subsection{Two-ion reorder energy barrier}\label{sec:reorderBarrier}
%%%%%%%%%%%%%%%%%%%%%%%%%%%%%%%%%%%%%%%%%%%%%%%%%%

The reorder probability depends on the operating parameters of the ion trap and the details of the laser cooling during the measurement.  Following \citet{Wubbena2012}, we write the total trap electric potential as
\begin{equation}\label{eq:trapPotential}
\Phi_T = \frac{V_0}{2} \cos (\Omega t) \frac{x^2 - y^2}{R^2}+U_0 \frac{z^2 - \alpha x^2 - (1 - \alpha) y^2}{d^2},
\end{equation}
\\
where $V_0/2$ and $\Omega$ are the amplitude and angular frequency of the voltage applied to the rf electrodes, $U_0$ is the voltage applied to the endcap electrodes, $R$ and $d$ are characteristic radial and axial dimensions of the trap, and $\alpha$ parameterizes the radial asymmetry of the static field.  The secular frequencies of a single ion are
\begin{align}
	\begin{split}
	\omega_z &= \sqrt{\frac{2 Q U_0}{m d^2}} \ , \\
	\omega_x &= \sqrt{\omega_p^2 - \alpha \omega_z^2} \ , ~ \textrm{and} \\
	\omega_y &= \sqrt{\omega_p^2 - (1 - \alpha) \omega_z^2} \ ,
	\end{split}
	\label{eq:trapfreqs}
\end{align}
where $m$ is the mass of the ion and
\begin{equation}
\omega_p = \frac{Q V_0}{\sqrt{2} \Omega m R^2} \ .
\label{eq:wp}
\end{equation}
We define the potential energy barrier, $\ereorder$, as the energy difference between the
ground-state axial crystal and the minimum energy configuration of two ions
confined to the radial plane.  Assuming that the coordinate system is defined
such that $\omega_x > \omega_y$, the potential energy barrier for changing the
order of a two-ion crystal is
\begin{widetext}
\begin{equation}
	\label{eq:reorderBarrier}
E_\textrm{reorder} = \frac{3}{4} \left( \frac{\sqrt{m} \omega_z Q^2}{2 \pi \epsilon_0} \right)^{2/3} \left[ \left( \frac{2 (\epsilon^2 + \alpha - 1)(\epsilon^2 + \mu (\alpha - 1))}{\epsilon^2 (\mu + 1) + 2 \mu (\alpha - 1)} \right)^{1/3} - 1 \right] \ ,
\end{equation}
\end{widetext}
where $\mu = m_2/m_1$ is the ratio of the two ion masses, $\epsilon = \omega_p/\omega_z$, and $\omega_z$ is the axial secular frequency of a single ion of mass $m=m_1$ in the trap.

In the absence of cooling, and assuming that the six secular frequencies are not rationally related, all collisions that deposit a kinetic energy greater than the reorder energy barrier will eventually cause the order of the two ion crystal to repeatedly change.
The order changes will continue until cooling is applied, at which time the order is fixed.  Assuming that neither the cooling process nor the trap potential introduces an order bias, the probability that the order after cooling is different from the order before the collision is 50\%.  Experimentally, however, it is more convenient to measure the reorder rate while applying continuous laser cooling.  In this case, collisions that deposit a kinetic energy slightly greater than the reorder energy barrier cause reordering with probability less than 50\%, with the exact probability depending on the operating parameters of the ion trap and the details of the laser cooling.

\begin{figure}
\begin{center}
	\includegraphics[width=1.0\columnwidth]{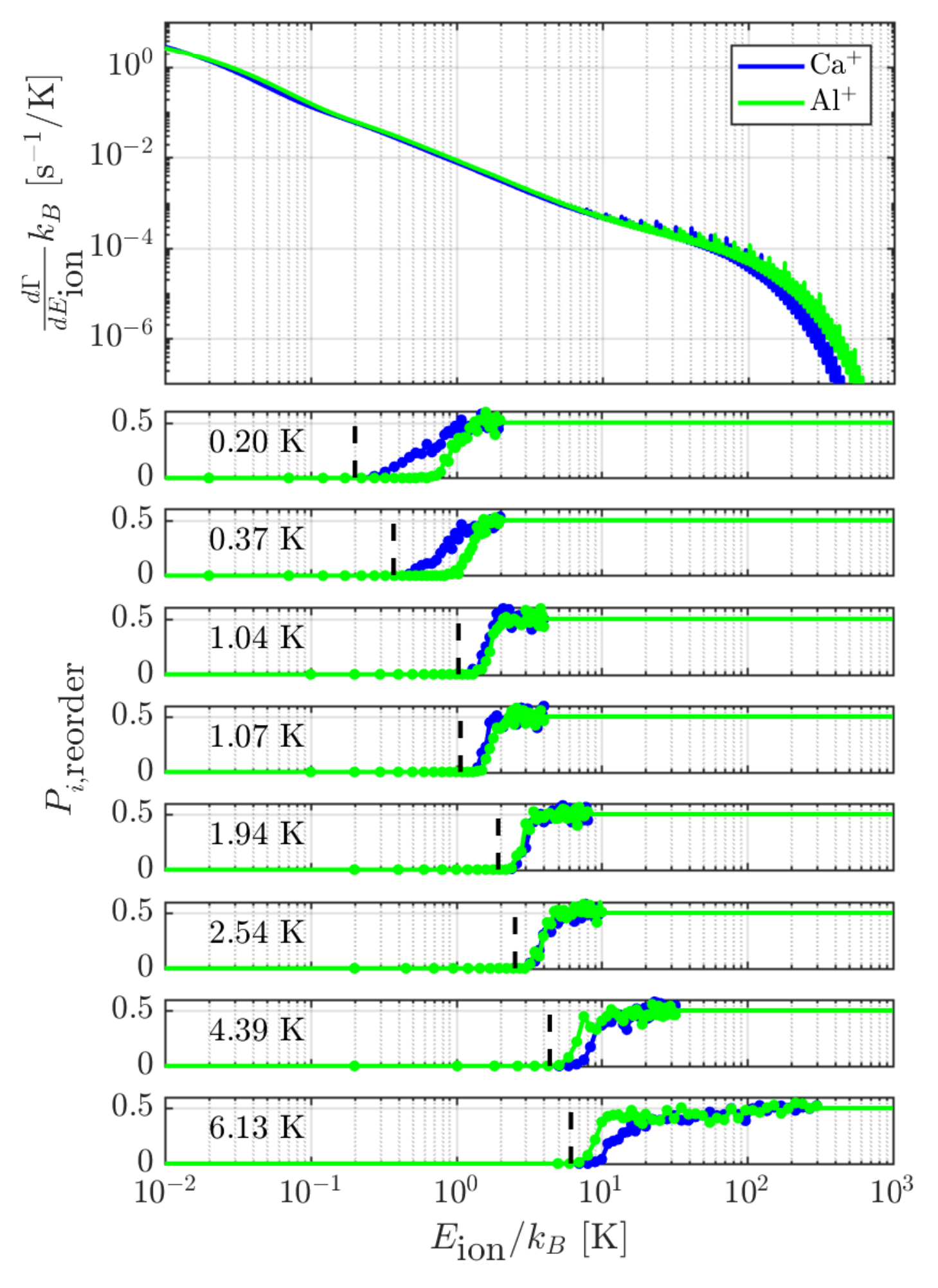}
	\caption{\label{fig:reorderProbability}(Color online) (top panel) Collision
	rate for each ion of a $\caalp$ two-ion crystal in 51.9~nPa of
	H$_2$ BG at 295~K. (other panels) Reorder probability calculated
	using the Monte-Carlo ion trajectory simulations.
	Each panel is for different trap parameters. The reorder barrier for each set
	of trap parameters is specified on the left side of the panel and is shown as
	a dashed vertical line. The laser cooling parameters used are a saturation
	parameter of $s = 300$ and a detuning of 80~MHz.
}
\end{center}
\end{figure}

%%%%%%%%%%%%%%%%%%%%%%%%%%%%%%%%%%%%%%%%%%%%%%%%%%
\subsection{Ion trajectory simulations}\label{sec:trajectorSims}
%%%%%%%%%%%%%%%%%%%%%%%%%%%%%%%%%%%%%%%%%%%%%%%%%%

We have calculated the reorder probability as a function of the ion energy
after the collision for several experimentally accessible reorder energy
barriers using Monte-Carlo simulations of the ion trajectories.  We calculate
the ion trajectories in the full time-dependent potential
(Eq.~\ref{eq:trapPotential}) using a custom sixth-order symplectic integrator
\cite{Yoshida1990}.  The integrator includes the nonlinear Coulomb interaction
between the two ions and a classical, linear drag force to approximate Doppler laser
cooling.  For each value of the energy transferred to an ion by a collision,
and for collisions with each of the two ions, we run the integrator for 256
randomly selected initial velocity directions.  We have checked the sensitivity
of the results to our treatment of the laser cooling by repeating the calculation
for saturation parameters that are a factor of 2 lower and higher than the
nominal value, and verified that the calculated BG pressure changes by
less than 1\%. The calculated reorder probabilities for the nominal
saturation parameter are shown in Fig.~\ref{fig:reorderProbability}. 

%%%%%%%%%%%%%%%%%%%%%%%%%%%%%%%%%%%%%%%%%%%%%%%%%%
\section{Two-ion crystal order detection}\label{sub:orderdetection}
%%%%%%%%%%%%%%%%%%%%%%%%%%%%%%%%%%%%%%%%%%%%%%%%%%

\begin{figure}%[htbp]
	\includegraphics[width=1.0\linewidth]{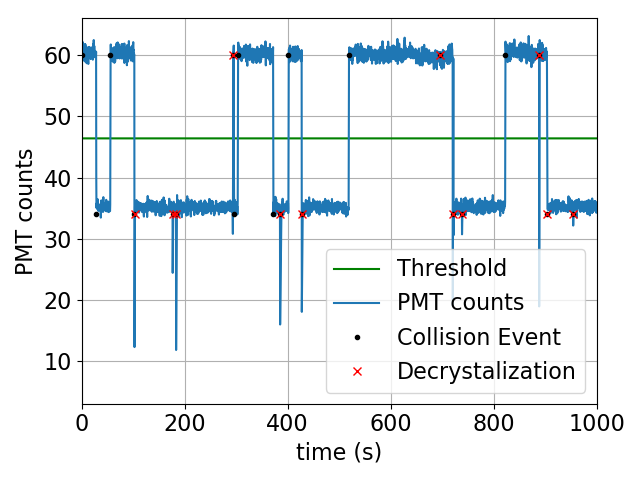}
	\caption{
		PMT counts (blue) observed during a typical reorder-rate measurement. A knife edge placed at the center of an optical image of the two-ion crystal causes a reduction in the measured count rate for the order in which $\ca$ is partially blocked. For this example, two-ion crystal order changes (black dots) are observed as sudden changes between 60 and 35~counts measured during a 800~$\mu$s measurement window. Fluorescence dropouts below 35~counts PMT counts arise due to energetic collisions that cause the two-ion crystal to dissolve (red crosses).
}
\label{fig:rawPMTcounts}
\end{figure}

The ion order is detected by spatially resolving the positions in the two-ion crystal.  A matched pair of aspheric lenses (NA = 0.38, 50 mm diameter) produce an image of the ions at an intermediate focus outside of the vacuum chamber.  A knife edge positioned close to the center of the two ions in this image plane blocks the $\ca$ ion fluorescence in one order while transmitting it in the other.  The intermediate image is relayed onto a PMT such that there are two well resolved fluorescence levels.  As an example, \autoref{fig:rawPMTcounts} shows a time series with an average of 60 counts in the bright order and 35 counts in the dark order for an 800~$\mu$s measurement window.

A reorder event causes a discrete jump in the observed fluorescence between the dark and bright levels, which we identify using a threshold.  Ideally, the number of reorder events, $\nreorder$, observed during a total observation time, $T$, gives the reorder rate $R = \nreorder/T$.  However, in the data, we also observe decrystallization events in which the collected fluorescence drops below the dark level and the ions cannot be efficiently laser cooled.  These events are detected in real time, triggering the control system to lower the radial confinement, which allows efficient laser cooling and recrystallization of the ions.  A bias in the ion order after recrystallization can give a bias in the inferred reorder rate.  To account for this, we also record the total number of decrystallization events, $\ndecryst$.  The unbiased reorder rate is then given by
\begin{equation}\label{eq:experreorderRate}
	R = \frac{\nreorder + \frac{\ndecryst}{2}}{T} \ .
\end{equation}
Here, $\nreorder$ includes only reorder events that are not associated with
decrystallization.  The factor of 1/2 in the second term of
Eq.~\ref{eq:experreorderRate} ensures that any collision event that imparts
enough kinetic energy to decrystallize the ions contributes with a probability
of 1/2 to reorder them. 

In addition to the recrystallization bias, we have observed a non-negligible probability of recording false decrystallization events when the ions are in the dark order.  For this reason, we estimate $\ndecryst$ by extrapolating from the number of decrystallization events beginning in the bright order, $N_{\mathrm{decryst},B}$, which has substantially less overlap with the count histogram for the decrystallized state.
We use $\ndecryst =  N_{\mathrm{decryst},B}(T/T_B)$,
where $T_B$ is the total time the ions spent in the bright order.  We assume Poisson statistics for $N_{\mathrm{reorder}}$ and $N_{\mathrm{decryst},B}$ to propagate the error for the error bars in Fig.~\ref{fig:reorderRate}.

%%%%%%%%%%%%%%%%%%%%%%%%%%%%%%%%%%%%%%%%%%%%%%%%%%
\section{Optical atomic clock collisional frequency shift and uncertainty}\label{sec:CollisionShift}
%%%%%%%%%%%%%%%%%%%%%%%%%%%%%%%%%%%%%%%%%%%%%%%%%%

\begin{figure}
\begin{center}
\includegraphics[width=1.0\columnwidth]{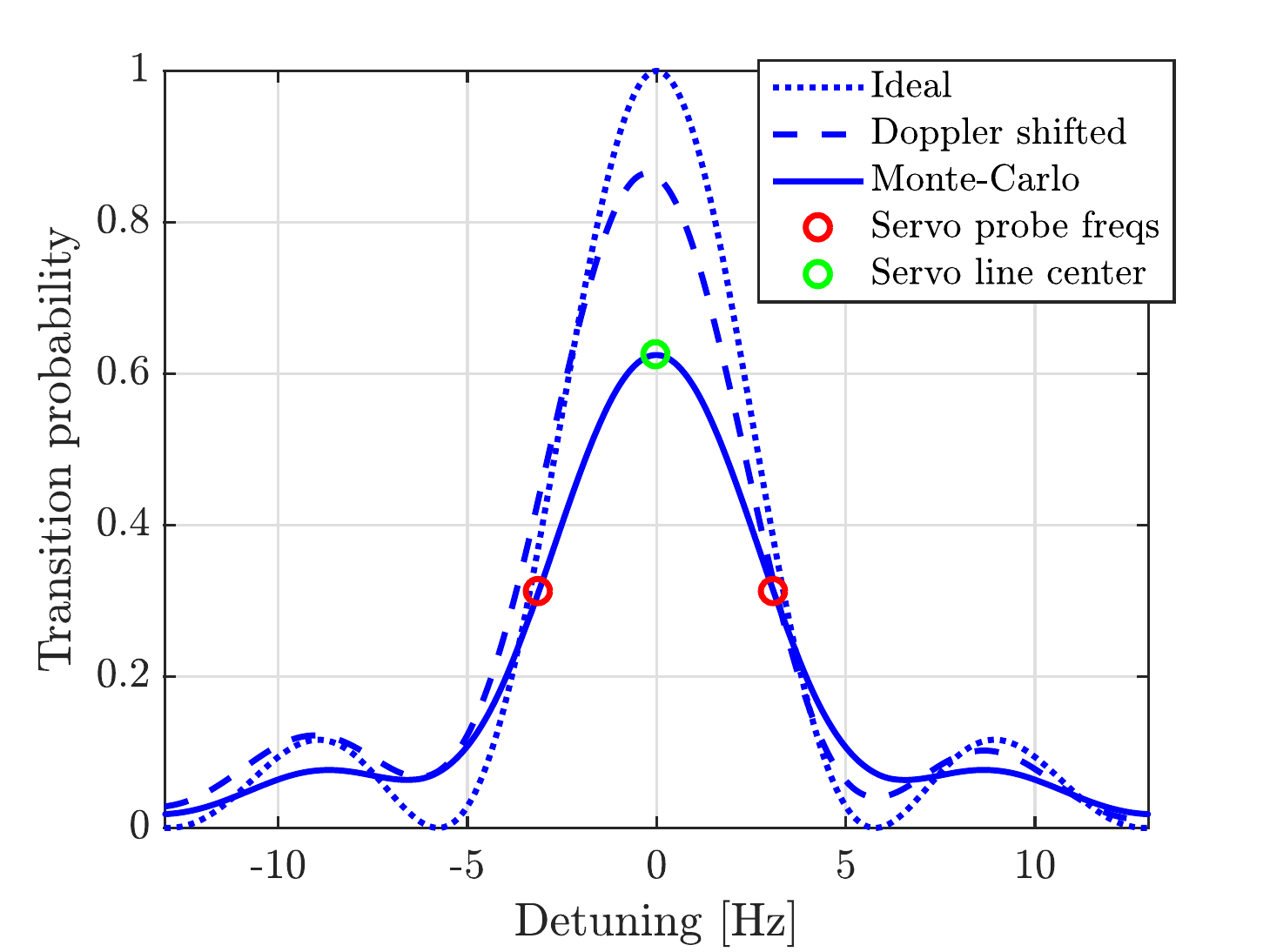}
\caption{\label{fig:transitionLineshape}(Color online) Rabi spectroscopy lineshape of a $^{25}$Mg$^+$ - $^{27}$Al$^+$ optical atomic clock operated with a 150~ms probe time.  Blue dotted line: spectroscopy lineshape for $^{27}$Al$^+$ at rest and without any BGCs.  Blue dashed line: Monte-Carlo spectroscopy lineshape including the time-dilation shift and phase shift due to collisions with 1.5~$\mu$Pa of H$_2$ BG at 294.15~K, but not including the Debye-Waller factors that suppress participation of high energy population.  This pressure is much higher than is achieved in experiments to exaggerate the asymmetry of the lineshape.  Blue solid line: Monte-Carlo spectroscopy lineshape including BGCs and Debye-Waller factors.  Red circles: frequencies that are probed during normal clock operation.  Green square: center frequency to which the optical clock output frequency is steered.}
\end{center}
\end{figure}

We calculate the background gas collision (BGC) shift for $^{27}$Al$^+$, $^{40}$Ca$^+$, $^{88}$Sr$^+$, and $^{171}$Yb$^+$ clocks using the Monte-Carlo model described in the main text.  The Monte-Carlo model input parameters and results for the four clocks are displayed in \autoref{tab:clockParameters}.  To illustrate the asymmetry of the clock transition lineshape due to collisions, Fig.~\ref{fig:transitionLineshape} shows the lineshape calculated by the Monte-Carlo model for input parameters selected to exaggerate the asymmetry.

In the following subsections, we present the details of the BGC shift and uncertainty calculation for the optical atomic clock described in \cite{Brewer2019}.  This clock is based on quantum-logic Rabi spectroscopy of the $^1$S$_0$ $\leftrightarrow$ $^3$P$_0$ transition in $^{27}$Al$^{+}$ using $^{25}$Mg$^{+}$ as the logic ion.  Unlike previous $^{27}$Al$^{+}$ clocks, this clock does not use any sympathetic cooling during the interrogation.  Furthermore, our Monte-Carlo model does not include sympathetic cooling during the interrogation, so it cannot be used to improve the collision uncertainty bound for the previous $^{27}$Al$^{+}$ clocks.  All of the calculations below assume a 150~ms probe time unless otherwise noted.

%%%%%%%%%%%%%%%%%%%%%%%%%%%%%%%%%%%%%%%%%%%%%%%%%%
\subsection{BG temperature}
%%%%%%%%%%%%%%%%%%%%%%%%%%%%%%%%%%%%%%%%%%%%%%%%%%

The temperature of the BG is evaluated using seven thermocouple sensors located on the trap wafer, in-vacuum support structure, and surrounding vacuum chamber.  We use the lowest and highest measured temperatures minus and plus the sensor uncertainty as the lower and upper bounds of the BG temperature to arrive at $294.15(2.70)$~K \cite{Brewer2019}.

%%%%%%%%%%%%%%%%%%%%%%%%%%%%%%%%%%%%%%%%%%%%%%%%%%
\subsection{BG pressure}
%%%%%%%%%%%%%%%%%%%%%%%%%%%%%%%%%%%%%%%%%%%%%%%%%%

The BG pressure is evaluated to be 38~nPa by measuring the rate of collisions that cause the two ions to swap positions.  Note that unlike the $^{40}$Ca$^+$ - $^{27}$Al$^+$ experiment, decrystallization is not observed in the $^{25}$Mg$^+$ - $^{27}$Al$^+$ experiment, so the analysis of this measurement is more straightforward.  There are several sources of uncertainty.  The 6~\% statistical uncertainty is limited by the 12~hour duration of the measurement.  We estimate the systematic uncertainty by considering the following approximations made in the model:
\begin{enumerate}
\item We calculate the differential cross section semiclassically based on the H$_2$ polarizibility and neglecting inelastic collisions.  For laser cooled $^{27}$Al$^{+}$ and $^{25}$Mg$^{+}$, the hyperfine and fine structure is energetically accessible in collisions with room temperature BG molecules.  For H$_2$ with a 1.8~THz rotational constant and 132~THz vibrational constant \cite{Huber2018}, the hyperfine and rotational structure is accessible but not the vibrational structure.  The contribution of resonant processes to momentum transfer from the BG to the ions is expected to be small, since the electric dipole matrix elements for these transitions are small.  Furthermore, we assume random scattering phases for $l < l_\textrm{crit}$ as described in Sec.~\ref{sec:differentialCrossSection}.  We assign a 50~\% systematic uncertainty to the BG pressure to account for these approximations, which we believe to be conservative.
\item We use the (experimentally measured) dc molecular orientation averaged H$_2$ polarizibility $\alpha = 0.80 \times 10^{-30}$~m$^3$ \cite{Landolt1951} to calculate the differential cross section.  The timescale for collisions is of order $b_\textrm{crit}/v$.  For the 1.8~km/s mean speed of H$_2$ at room temperature, $b_\textrm{crit} = 0.5$~nm and the collision timescale is 1/(4~THz).  We estimate the systematic uncertainty associated with this approximation to be 10~\% by calculating the pressure using the minimum and maximum of the parallel and perpendicular polarizibility at DC and 1~PHz ($0.68 \times 10^{-30}$~m$^3$ and $1.04 \times 10^{-30}$~m$^3$, calculated by \citet{Akindinova2009}).
\item The Monte-Carlo trajectory simulations use a linear drag model of laser cooling which overestimates the cooling force for high ion temperatures.  We estimate the uncertainty associated with this approximation to be 1.0~\% by running the trajectory simulations for cooling laser saturation parameters a factor of 2 higher and lower than the experimental value and calculating the pressure.  This estimate also captures the convergence uncertainty of the trajectory simulations.
\end{enumerate}
Finally, there is a 2.1~\% convergence uncertainty associated with the calculation of $\dgde(\eion)$ and a 0.7~\% uncertainty due to the temperature uncertainty.  Adding all of these uncertainties in quadrature, we arrive at a total uncertainty of 51~\% and a pressure of $38(19)$~nPa.  The uncertainty of this measurement could be significantly reduced by performing more accurate quantum scattering calculations and increasing the measurement time.

%%%%%%%%%%%%%%%%%%%%%%%%%%%%%%%%%%%%%%%%%%%%%%%%%%
\subsection{Background gas collision shift}\label{sec:BGCShift}
%%%%%%%%%%%%%%%%%%%%%%%%%%%%%%%%%%%%%%%%%%%%%%%%%%

We calculate the BGC shift using the Monte-Carlo model described in the main text with $10^8$ trials.  The time dilation component of the shift is $-0.60 \times 10^{-19}$ and the phase shift component is $\pm 2.32 \times 10^{-19}$.  At $10^8$ trials, the Monte-Carlo convergence uncertainty is approximately $10^{-21}$.  The convergence uncertainty associated with the calculation of $\dgde(\eion)$ is also approximately $10^{-21}$.  As the phase shift component is conservatively bounded by using the worst case $\pm \pi/2$ values for the collisional phase shift, we do not add any systematic uncertainty to this component.  Note that any resonant collisional phase shift processes are covered by this worst case bound, and that calculations of the collisional phase shift for other collision partners suggest that the phase shift component of the collisional frequency shift is likely much smaller \cite{Vutha2017}.  We estimate the systematic uncertainty of the time-dilation component by considering the following approximations:
\begin{enumerate}
\item Approximations 1 and 2 for the BG pressure evaluation described above effect the momentum transfer to the ions, and thus the suppression of the BGC shift due to the low contribution of collision events to the spectroscopy signal.  We assume that the uncertainty due to these approximations is captured by propogating the pressure uncertainty through the collision shift model, which yields an uncertainty on the time-dilation component of $\pm 0.32 \times 10^{-19}$.
\item While the reduction of the carrier Rabi frequency due to secular motion is calculated to arbitrary order in the Lamb-Dicke parameter following \citet{Wineland1998}, we do not include any reduction of the carrier Rabi frequency due to micromotion.  The reduction in Rabi frequency due to excess micromotion is compensated experimentally by increasing the laser intensity to produce a $\pi$ pulse.  The reduction in Rabi frequency due to intrinsic micromotion is a factor of $q/8$ smaller than that due to secular motion, where $q$ is the Mathieu parameter.  While this approximation results in a small change in the total Rabi frequency reduction, it may cause us to underestimate the large suppression of the time-dilation component due to the Debye-Waller effect.  Thus we expand the uncertainty range of the time-dilation shift to include zero, resulting in a bound of $-0.60^{+0.60}_{-0.32} \times 10^{-19}$
\item Our calculation of the spectroscopy lineshape does not include any decoherence other than that due to collisions.  As the experimentally measured spectroscopy lineshape has $\gtrsim 70$~\% contrast and a Fourier-limited width \cite{Brewer2019}, we do not believe this approximation has a significant effect on the calculated BGC shift.
\end{enumerate}
Adding the uncertainty of the time-dilation and phase shift components in quadrature, we arrive at a total BGC shift of $-0.60^{+2.40}_{-2.34} \times 10^{-19}$ for a probe duration of 150~ms.  In the main text, we take the maximum of the upper and lower uncertainties to arrive at $-0.6(2.4)\times 10^{-19}$.  The uncertainty of this calculation could be reduced by performing more accurate quantum scattering and frequency shift calculations \cite{Vutha2017, Vutha2018}, although a significant reduction would require accurate knowledge of the partial pressures of each BG species.

%%%%%%%%%%%%%%%%%%%%%%%%%%%%%%%%%%%%%%%%%%%%%%%%%%
\subsection{Other considerations}\label{sec:OtherConsiderations}
%%%%%%%%%%%%%%%%%%%%%%%%%%%%%%%%%%%%%%%%%%%%%%%%%%

In all of the above, we have assumed that the background gas is dominantly H$_2$, which is typically the case for room temperature UHV systems that have undergone a high temperature bake.  The stainless steel vacuum parts used for this clock (excluding the windows and feedthroughs) were prebaked under vacuum at 400~C for 7 days, and the assembled chamber was baked at 145~C for 7 days.  To gauge whether small amounts of other BG species have a significant impact on the pressure and BGC shift results, we have calculated the pressure and BGC shift for a variety of other background gases common in UHV chambers, with the results shown in Tab.~\ref{tab:collisionShiftVsSpecies}.  The calculated BGC shifts under the assumption that 100~\% of the BG is a species other than  H$_2$ are not drastically different than that for H$_2$, and the partial pressures of other BG species in our vacuum chamber are expected to be much lower than that of H$_2$, so we do not increase the BGC shift uncertainty to account for small amounts of other BG species which may be present.

\begin{table*}
\caption{\label{tab:collisionShiftVsSpecies} Pressure and collision shift of the $^{27}$Al$^+$ clock described in \cite{Brewer2019} calculated assuming that 100~\% of the BG is the species listed in the left column.}
\begin{ruledtabular}
\begin{tabular}{llllll}
BG species				& Polarizibility [$10^{-30}$~m$^3$]	& Pressure [nPa]		& \multicolumn{3}{l}{Fractional frequency shift [$10^{-19}$]} \\
						&									&					& Time-dilation			& Phase shift	& Total \\
\hline
H$_2$\rule{0pt}{3ex}	& 0.80 \cite{Landolt1951}				& $38 \pm 19$		& $-0.6^{+0.6}_{-0.3}$	& $\pm 2.3$		& $-0.6 \pm 2.4$ \\
He\rule{0pt}{3ex}		& 0.21 \cite{Lach2004}				& $88 \pm 45$		& $-0.4^{+0.4}_{-0.2}$	& $\pm 1.1$		& $-0.4 \pm 1.2$ \\
CH$_4$\rule{0pt}{3ex}	& 2.43 \cite{Maroulis1994}			& $36 \pm 18$		& $-0.8^{+0.8}_{-0.4}$	& $\pm 3.0$		& $-0.8 \pm 3.1$ \\
CO\rule{0pt}{3ex}		& 1.97 \cite{Bridge1966}				& $45 \pm 23$		& $-0.8^{+0.8}_{-0.4}$	& $\pm 3.0$		& $-0.8 \pm 3.1$ \\
N$_2$\rule{0pt}{3ex}	& 1.76 \cite{Bridge1966}				& $48 \pm 25$		& $-0.8^{+0.8}_{-0.4}$	& $\pm 2.8$		& $-0.8 \pm 2.9$ \\
O$_2$\rule{0pt}{3ex}	& 1.59 \cite{Bridge1966}				& $52 \pm 27$		& $-0.8^{+0.8}_{-0.4}$	& $\pm 2.8$		& $-0.8 \pm 2.9$ \\
CO$_2$\rule{0pt}{3ex}	& 2.63 \cite{Bridge1966}				& $44 \pm 23$		& $-0.8^{+0.8}_{-0.4}$	& $\pm 3.8$		& $-0.8 \pm 3.9$
\end{tabular}
\end{ruledtabular}
\end{table*}

We do not include an atomic phase shift due to glancing collisions in our simulations.  We have estimated the magnitude of the contribution of glancing collision atomic phase shifts to the BGC shift using the following calculation.  For a given BG velocity and impact parameter, the classical trajectory of the BG and the electric field produced at the position of $^{27}$Al$^+$ by the polarized H$_2$ molecule is calculated.  Using the experimentally measured differential polarizibility of the $^{27}$Al$^+$ clock transition \cite{Brewer2019}, the time dependent Stark shift is integrated to obtain the phase shift of $^{27}$Al$^+$ resulting from the collision.  This phase shift is averaged over a Boltzmann distribution of BG speeds and impact parameters and divided by the collision rate to obtain a frequency shift of order $10^{-22}$, which is negligible.

%%%%%%%%%%%%%%%%%%%%%%%%%%%%%%%%%%%%%%%%%%%%%%%%%%
\bibliography{collision_uncertainty}
%%%%%%%%%%%%%%%%%%%%%%%%%%%%%%%%%%%%%%%%%%%%%%%%%%

\end{document}